\documentclass[preprint2,longabstract]{aastex62}

\usepackage{natbib}

\def\msun{M$_{\odot}$}
\def\rsun{R$_{\odot}$}

\def\mdot{$\dot M$}

\def\it{\sl}
\def\degs{\ifmmode ^{\circ}\else$^{\circ}$\fi}
\def\amin{\ifmmode ^{\prime}\else$^{\prime}$\fi}
\def\asec{\ifmmode ^{\prime\prime}\else$^{\prime\prime}$\fi}
\def\farcs{\hbox{$.\!\!^{\prime\prime}$}}  % Fractions of arcseconds
\def\degs{\ifmmode ^{\circ}\else$^{\circ}$\fi}
\def\amin{\ifmmode ^{\prime}\else$^{\prime}$\fi}

\def\eqalign#1{\null\,\vcenter{\openup1\jot \m@th
   \ialign{\strut\hfil$\displaystyle{##}$&$\displaystyle{{}##}$\hfil
   \crcr#1\crcr}}\,}
\usepackage{color}
\usepackage{enumerate}

\newcommand{\myr}{\mbox {~${\rm M_{\odot}~yr^{-1}}$}}

\newcommand{\sgr}{V1082\,Sgr}
\def\apgt{\ {\raise-.5ex\hbox{$\buildrel>\over\sim$}}\ }
\def\aplt{\ {\raise-.5ex\hbox{$\buildrel<\over\sim$}}\ }

\newcommand{\kms}{km\,s$^{-1}$}

\begin{document}

\title{Quest for the Donor Star in the Magnetic Precataclysmic Variable V1082 Sgr}

\author{G. Tovmassian}
\affiliation{Instituto de Astronom\'{\i}a, Universidad Nacional Aut\'onoma de M\'exico, Apartado Postal 877, Ensenada, Baja California, 22800, M\'exico}

\author{J. F. Gonz\'alez}
\affiliation{Universidad Nacional de San Juan, Av. J. I. de la Roza 590 oeste, 5400 Rivadavia, San Juan, Argentina}
\affiliation{ICATE, CONICET, Av. Espa\~na 1512 sur, J5402DSP San Juan, Argentina}
 
\author{M. - S. Hern\'andez}
\affiliation{Instituto de F\'{i}sica y Astronom\'{i}a, Facultad de Ciencias, Universidad de Valpara\'{i}so, Av. Gran Breta\~{n}a 1111 Valpara\'{i}so, Chile}

\author{D. Gonz\'alez--Buitrago}
\affiliation{Department of Physics and Astronomy, 4129 Frederick Reines Hall, University of California, Irvine, CA 92697-4575, USA}

\author{S. Zharikov}
\affiliation{Instituto de Astronom\'{\i}a, Universidad Nacional Aut\'onoma de M\'exico, Apartado Postal 877, Ensenada, Baja California, 22800 M\'exico}

\author{J. V. Hern\'andez Santisteban}
\affiliation{Anton Pannekoek Institute for Astronomy, University of Amsterdam, Science Park 904, NL--1098 XH Amsterdam, The Netherlands}

\email{gag@astro.unam.mx}

\begin{abstract} 
We obtained high-resolution spectra and multicolor photometry of  V1082\,Sgr to study the donor star in this 
20.8\,hr orbital period binary, which is assumed to be a detached system.  We measured the rotational velocity  ($v\sin i =26.5\pm 2.0$ \kms),
which, coupled with the constraints on the white dwarf mass from the X-ray spectroscopy, leads to the conclusion 
that the donor star barely fills 70\%  of its corresponding Roche lobe radius. It appears to be a slightly evolved K2-type  star.
This conclusion was further supported 
by a recently published distance to the binary system measured by the {\it Gaia} mission.
At the same time, it  becomes difficult  to explain a very high ($> 10^{-9}$\myr)  mass 
transfer  and mass accretion rate in a detached  binary via stellar wind and magnetic coupling. 
\end{abstract}

\keywords{ binaries: close -- novae, cataclysmic variables,  --  stars: individual (V1082\,Sgr) -- star: rotation -- stars: winds, outflows } 
%\maketitle

\section{Introduction}

The term magnetic pre-polars was coined by \citet{2009A&A...500..867S}, when it was recognized that some magnetic white dwarfs (MWDs) are accreting matter from the secondary,
an active late-type main-sequence star underfilling its Roche lobe. The spectra of these systems show strong cyclotron harmonics 
in the form of wide humps superimposed on the WD+M-dwarf stellar continuum. Accretion onto the MWD primary is
of the order of $10^{-14} - 10^{-13}$\,\myr, comparable  to what is expected from the wind of a chromospherically active companion 
star \citep{2002ASPC..261..102S}.  \citet{2015SSRv..191..111F} listed 10 such systems, all of which contain
M dwarfs as a donor star.  It is natural to assume that there also should exist pre-polars with donor stars of earlier spectral types. 
It is not clear how justified the term pre-polars would be for wider systems with earlier-type companions. %, but let's expand here the definition of  pre-polars.
 But let us assume that  pre-polars are detached binaries consisting of a strongly MWD and any late-type zero-age main-sequence 
 (or nearly so)  magnetically active star,
in which the magnetic fields are coupled, forcing synchronous rotation of the components and driving mass transfer. Observationally, pre-polars with
earlier-type companions might manifest themselves differently than those containing an M dwarf. 

 \citet{2016ApJ...819...75T,2017ASPC..509..489T} proposed two candidates for pre-polars with early-K companions. One of them is \sgr, remarkable for its cyclical accretion activity 
 and low-luminosity episodes during which the K star is the predominant source of light.
\sgr\ is a prominent X-ray source. \citet{2013MNRAS.435.2822B} studied it with several available X-ray telescopes and concluded 
that it is a highly variable X-ray source  with a spectrum matching those of magnetic cataclysmic variables (CVs). 
They identify a small  X-ray-emitting region where  the plasma has typical temperatures achieved in a
magnetically confined accretion flow. Using the model of  \citet{2005A&A...435..191S}, the mass of the magnetically accreting WD was estimated 
to be $M_{\mathrm {wd}} = 0.64\pm0.04$\,\msun. From the derived WD mass and radius ($8.3\times10^8$\,cm), they deduced a mass accretion rate of 
 \mdot = $2-4\times10^{-9}$\,\msun\,yr$^{-1}$\  for a distance of 730 pc and 1.15 kpc, respectively.

We conducted  high-resolution spectroscopy  accompanied by parallel multiband photometry to define the parameters 
of the donor star and the binary as a whole. This study comes on the heels of 
observations  of the object by the {\sl Kepler K2} mission. Results of  80 days long of continuous, time-resolved photometry 
of the object are analyzed by \citet[][hereafter Paper I ]{2018ApJ...863...47T}. Relevant to 
this follow-up article  is the detection of  the orbital period in the {\sl K2} light curve. %Particularly, it is clearly visible when \sgr\ 
%s in the deep low state and the light comes predominantly from the donor star. 
In \sgr,  there are  deep low states when the light curve appears to be dominated by the donor star. Paper I concludes that such  a light curve cannot be
produced by an ellipsoidally deformed star because it would create two dips per orbit,  and hence the K star in this binary is not filling its Roche lobe.
Instead, one dip per orbit has been observed, which was interpreted in Paper I as presence of  a  spot  (cool, hot, or a combination of both) on the surface
of the donor star. 

In this paper, we approach the same problem from a different point of view to confirm that \sgr\ is indeed a detached binary.
In Section~\ref{sec:obs}, we describe our observations of \sgr\ and the
corresponding data reduction.  In Section~\ref{abs}  we present
the analysis of a complex of absorption lines and the measurements of  radial velocities (RVs). The deduction of rotational velocity
is in Section~\ref{rotation}  and the binary system parameters  are presented in Section~\ref{param}.
In Section \ref{emlines}, we review the new information gathered from the emission-line profiles. 
We provide  a discussion of the obtained results and their application  in  Section~\ref{conclud}.

\section{Observations}
\label{sec:obs}

The high-resolution spectroscopic observations  of \sgr\ were obtained using the echelle REOSC
spectrograph \citep{Levine1995}    at  the 2.1\,m telescope of
the Observatorio Astron\'omico Nacional at San Pedro M\'artir (OAN
SPM)\footnote{http://www.astrossp.unam.mx}, Mexico.  The echelle
spectrograph provides spectra %spread over 27 orders, 
covering the  $\sim$3500--7105~\AA\ range with a spectral resolving power of
R$\approx$18,000. A total of 42 echelle spectra were obtained during
11 consecutive nights in 2017 July. A Th-Ar lamp was
used for wavelength calibration. The spectra were reduced using the {\it
echelle} package in {\sc iraf}\footnote{IRAF is distributed by the National Optical Astronomy Observatories, which
are operated by the Association of Universities for Research in Astronomy, Inc., under cooperative agreement with the National Science Foundation.}.  
Standard procedures, including bias subtraction, cosmic-ray removal, and wavelength  calibration were carried out
using  the corresponding tasks in {\sc iraf}. The flux calibration is notoriously difficult with echelle spectra, which we did not attempt. 
We merged all orders
after normalizing them to the continuum. In the overlapping regions,  spectra were weighted according to their signal level, and the ends 
of the orders were apodized with a cosine bell to prevent discontinuities.  These merged spectra were used for the spectral 
class and rotational velocity determination.
There is a  problem with the focus of the REOSC spectrograph,  originally  designed for  photographic plates  
but now  modified for a CCD camera that provides a larger field of view, and 
the focus deteriorates  toward the ends of  the spectral orders.

A number of K stars of different spectral and luminosity classes of known rotational velocities used
routinely as standards  \citep{1997PASP..109..514F} were observed along the object.  
HD\,182488 was primarily used for rotational velocity measurements, 
together with HD\,166620 for
RV measurements. 
Several others  of earlier and later spectral types, as well as luminosity class IV, were observed and used for spectral class
determination.
The log of observations is given in  Table~\ref{tab:log}.

%+================++++++++++++++++++==================
\begin{table}[t]
\begin{center}
\footnotesize
\caption{\small  Log of spectroscopic observations  obtained with the 2.1m telescope and the echelle spectrograph.}
\begin{tabular}{lclc}
\hline\hline

 Date  UT                           &JD        & $t_{\mathrm{exp}}$ & Number \\
  yyyy\,mm\,dd                                   &2,450,000+                      &    (s)    &  of Spectra  \\\hline
2017 07 07      	    &7941&      1200                       & 1 \\
2017 07 08               &7942 &      1200                       & 2 \\
2017 07 10               &7944&      1800                       & 1 \\
2017 07 11               &7945&      1200                       & 2 \\
2017 07 12               &7946&      1200                       & 2 \\
2017 07 13               &7946&      1200                      & 2 \\
2017 07 14               &7948&      1200                      & 8 \\
2017 07 15      	    &7949&      1200                       & 7 \\
2017 07 16               &7950 &      1800                       & 8 \\
2017 07 17               &7951&      1800                       & 7 \\
2017 07 18               &7952&      1800                      & 2 \\\hline\hline
\end{tabular}
\label{tab:log}
\end{center}
\begin{tabular}{l}
\end{tabular}
\end{table}

Multicolor photometric observations were obtained with the Reionization and Transients InfraRed (RATIR), a 
simultaneous six-filter imaging camera ($r, i, Z, Y, J$ and $H$\ bands), 
mounted on a  Harold L. Johnson 1.5\,m telescope at OAN \citep{2012SPIE.8446E..10B,2012SPIE.8444E..5LW}. 
It operates in robotic mode and is available
in the absence of gamma ray burst alerts. We asked for sequences of multicolor exposures prior, during, and after the spectral observations. 
The telescope is not designed
for prolonged monitoring or long exposures, and the guiding is very poor. Hence, the reduction of the data was very arduous 
and required one-by-one inspection
of all images, a  large fraction of which turned out to be worthless as a result of bad pointing or guiding. However, we could use the 
good data to produce  decent light
curves in the $V$, $J$ and $H$\  bands, and get a good idea of the luminosity state of the object during spectroscopic observations. 
The images were processed using an automatic pipeline package for bias subtraction, flat-fielding,  and cosmic-ray removal. 
The reduction pipeline also permits sky subtraction when necessary and  
astrometric alignment of images. For the latter,  \url{astrometry.net}\ 
software was used.

After these preliminary steps, we measured the magnitudes of the object and of several similar and brighter comparison stars 
found in the field  using the {\sc iraf} task  {\it apphot} within the {\sc daophot} package. %The aperture was selected to minimize the errors
%and to underline the fast variability. 
We used a circular aperture with 5\farcs5 radius for the object and comparison stars.  
The object magnitudes were then determined in a differential photometry with  comparison stars of known magnitudes. 
We checked  the comparison stars against each other to ensure that none of them were variable and determined errors of measurements 
as a standard deviation.

\begin{figure}[t]
\centering
\includegraphics[width=8.0cm, bb = 20 160 580 700, clip=]{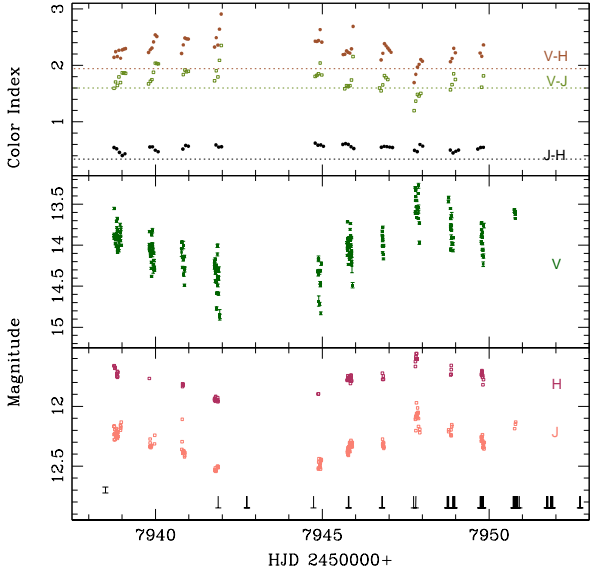}
\caption{Light curve of \sgr\ in three  bands during the  spectroscopic observations. 
In the bottom panel,  measurements  of IR  observations are plotted. Epochs of spectral exposures are marked in the bottom of the panel.
In the left bottom left corner,  the photometric error bar is indicated. The $V$-band light curve is presented in the middle panel, 
In the top panel, color indexes are presented. All photometric bands and color indexes are plotted in distinct colors 
and indicated on the right sides of the light curves. The horizontal dotted lines 
in the top panel indicate colors corresponding to a K2 main-sequence star. 
%The statistical measurement error  derived as a $3\sigma$ strip width 
%of similar brightness comparison star is marked by a bar in the right lower corner of the plot. 
%The magnitudes are instrumental  and are placed to overlap, do not reflect real colors or actual brightness 
}\label{lc}
\end{figure}

\section{Analysis of Absorption Lines}
\label{abs}

The spectrum of \sgr\ shows  absorption lines from the K donor star throughout the wavelength range covered by our data set. 
Often the continuum is contaminated by the accretion-fueled radiation taking place in this system, as well as by emission 
lines of hydrogen and helium.  However, as was already mentioned above, \sgr\ 
undergoes low-state episodes, when the contribution of the donor star is overwhelming.  There appear to be brief intervals when the 
accretion stops completely,  exposing  pure K-type spectrum in the optical range. 
\citet{2016ApJ...819...75T}  demonstrated examples of such spectra and suggested a K2 spectral class. In the new, 
high-resolution observations, we caught the system in the deep minimum, but it was too faint to get a good signal-to-noise 
(S/N) ratio spectrum with the echelle spectrograph.  
However, for the analysis of absorption lines and their profiles,
the spectra obtained in a higher-luminosity state are fine, since absorption lines are better exposed on the 
background of the elevated continuum.  In Figure\,\ref{lc} the spectral exposures are marked at the bottom of the 
light curve obtained by RATIR in near-IR filters.  The brightness of the system
constantly changes with the S/N at the continuum, reaching $\sim20$ in the single brightest spectra 
and  about 2 in the faintest phase. 

\subsection{Spectral type and RVs}

\begin{figure}[t]
\centering
\includegraphics[width=8.0cm, bb = 10 30 540 710, clip=]{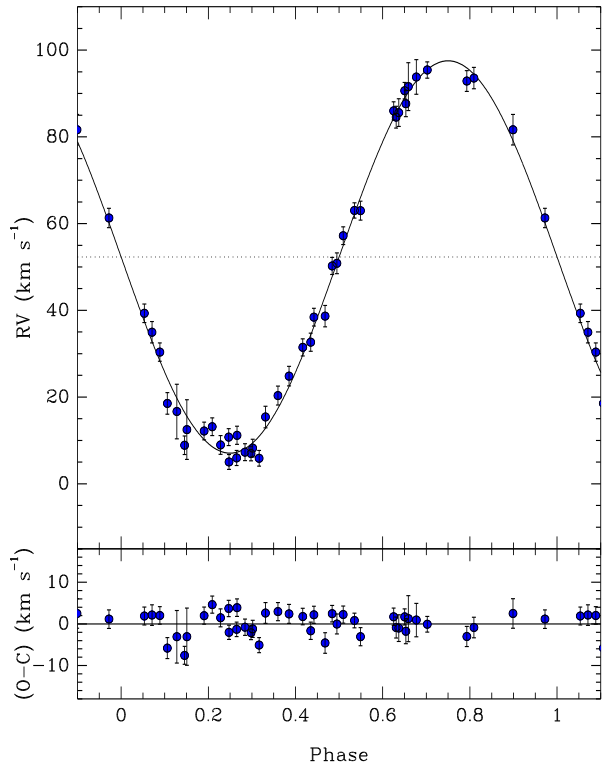}
\caption{The RV curve of the donor star measured by cross-correlation with the synthetic spectrum. Systemic velocity is marked with a dotted line.
%The red points are measurements against the HD\,182488, the blue points 
%corresponding measurements for the HD\,166620. The red and blue curves are sinusoidal fits to the corresponding points. The black curve is the combined average. 
The parameters of the
average fit are presented in Table\,\ref{tab:param}. }\label{rvcurve}
\end{figure}

%\subsection{Radial velocity and orbital analysis}

Measurements of  RVs %in the high-resolution spectra 
improved greatly compared to previously 
available data \citep{2010PASP..122.1285T,2016ApJ...819...75T}.
We cross-correlated  the ranges of 
spectra containing multiple strong absorption features with  reference spectra of standard stars. 
We used spectra of HD\,182488 (K0\,V), HD\,166620(K2\,V),  which were observed with the same instrumental settings together with the object, 
and a synthetic  spectrum of $T_\mathrm{eff}=5000$ K taken from the {\sc bluered} database \citep{2008A&A...485..823B}.
We refined the orbital period  by analyzing the new RV measurements in combination with the previous data \citep{2016ApJ...819...75T} 
for a longer time base.
We fitted the  RV curve using only data from high-resolution observations by fixing the period obtained from a larger database.
The differences in the results obtained using different templates are negligible. Formal  errors of  measurements are smaller 
by a factor of $\sim 2$ in the case of the synthetic spectrum, although the residuals are similar for either case (only $\sim 10$\% smaller). 
The RV measurements are included in electronic tables accompanying the paper. %listed in Table~\ref{tab:rvs}.  

The results of all measurements are summed up 
in Fig.~\ref{rvcurve} and Table~\ref{tab:param}.   
The parameters listed in the table are  
time of primary conjunction, center-of-mass velocity, velocity amplitude, orbital period, and standard deviation of residuals. 
The orbit was assumed to be circular.

\begin{table}[h]
\caption{Parameters of the RV Curve.}\label{tab:param}
\centering
\vspace{2 mm}
\begin{tabular}{lcr@{$\pm$}l}\hline\hline
$T_\mathrm{0}$ & HJD	& 2,457,939.161&	0.002\\
$V_\gamma$	&\kms	&51.8		&0.6 \\
{$K_d$} & \kms & 45.3 & 0.7\\
$P$&days	&0.867525	 &0.000015\\ 
{$\sigma$}  & {\kms} & \multicolumn{2}{c}{2.8} \\ \hline\hline
\end{tabular}
\label{tab:param}
\end{table}

\begin{figure*}[t]
\centering
\includegraphics[width=10.0cm,height=15cm,angle=-90, bb = 25 40 540 740, clip=]{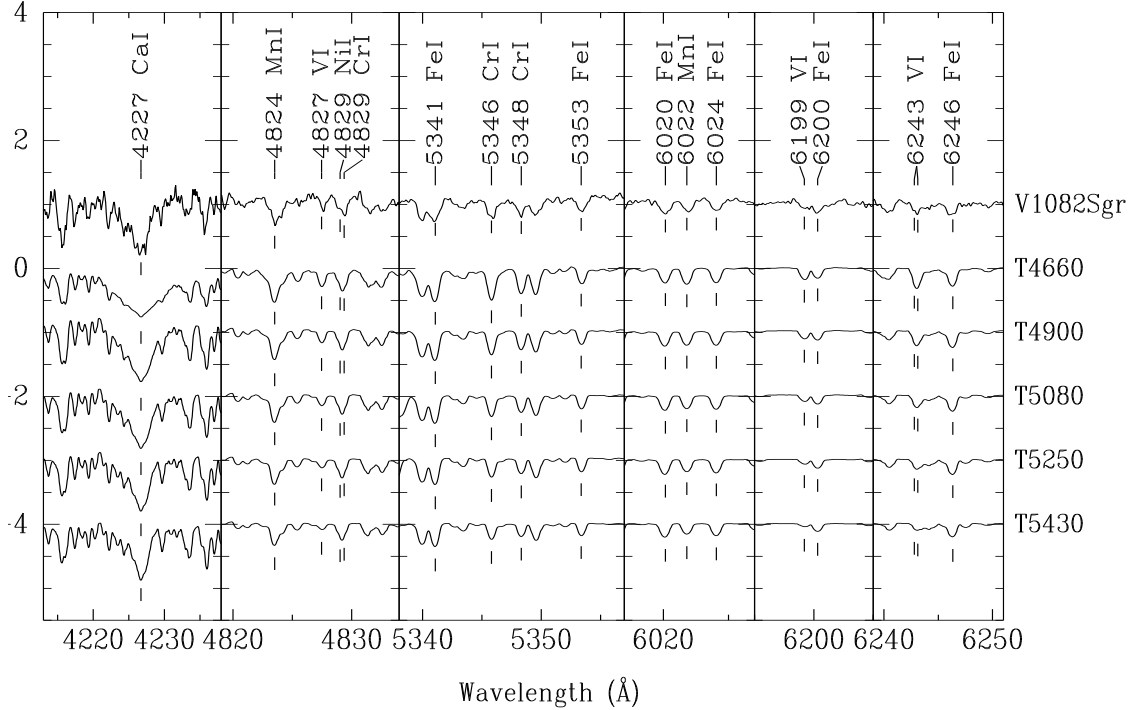}
\caption{Comparison of the spectral morphology of V1082\,Sgr with reference spectra of different spectral types.
HD\,182488 (K0\,V), HD\,142980 (K1\,IV),  HD\,115404 and HD\,166620 (averaged as K2\,V), and HD\,219134 (K3\,V) 
were observed along with \sgr\ during the same period of observations.}\label{sptype}
\end{figure*}

In order to produce a high-S/N  spectrum of the donor star, we proceeded as follows. First, we corrected the 
Doppler displacement of each merged spectrum 
according to the measured RV of the absorption lines. 
Then, we estimated the light contribution of the donor star by comparing the spectral-line intensities with those of a standard star. 
More precisely, we calculated the multiplying factor minimizing the difference between the object spectrum and the scaled reference spectrum 
in four small selected spectral regions and then averaged the four values into a single scale factor for each spectrum.
Typically, the scale factors range from 0.25 to 0.6 
depending on the luminosity state of the object, as reflected in the light curve presented in Figure\,\ref{lc}.
%with an average of 0.38.

Then the lines of the donor star were removed from each spectrum to measure the noise and identify 
deviant pixels to be removed. 
The individual spectra corrected for RV, after being scaled and cleaned, were combined to produce an average spectrum.  
In this calculation, we used optimal weights that  were calculated from the scale factors and the measured noise. 
%Unfortunately, the spectra at the minimum brightness of the object in which the contribution of the donor star are largest have small weights due to the increased noise.
%\fed{\em To me this last sentence is not entirely true, at least in theory. The effective signal-to-noise (=intensity of donor-star lines / noise) is expected to get worse as the light contamination (contribution of the companion) increases, mainly due to the shot noise of that extra light. }

We compared the combined averaged spectrum with reference spectra of different spectral types.
We selected from the Simbad database\footnote{http://simbad.u-strasbg.fr/simbad/sim-fsam.} stars in the temperature range 4500--5500 K, 
with solar metallicity (0.0$\pm$0.1 dex), and surface
gravity corresponding to main-sequence stars ($\log g$ = 4.3--4.6).
Then, we downloaded from the ESO archive\footnote{\url{http://archive.eso.org/wdb/wdb/adp/phase3_spectral/form.}}
 high S/N ($> 200$) spectra of these objects taken with the HARPS spectrograph.

The selected stars are listed in Table\,\ref{spstnd}.

\begin{table}[h]
\caption{List of Stars Used to Produce Templates}
\begin{center}
\begin{tabular}{lcccc}
\hline
star  &      $T_{\mathrm {eff}}$  &  $\log g$  &  Met &   SpType \\
\hline
HD 131977   & 4669  &  4.29  &  -0.04  &  K4\,V \\
HD 160346  &  4871  & 4.51   &  +0.00  &  K3\,V \\
HD 23356    &  4924  &  4.55  &  -0.08  &   K2.5\,V \\
HD 192310   & 5077   &  4.50  &   +0.04 &   K2\,V \\
HD 22049     & 5090   &  4.55  &  -0.07  &  K2\,V \\
HD 149661   &  5254   &  4.54 &   +0.03 &   K1\,V \\
HD 165341    & 5260   & 4.51  &  +0.00  &  K0\,V \\
HD 69830      & 5396   & 4.47  &  -0.05  &  G8\,V \\
HD 152391    & 5467   & 4.49  &  -0.02  &  G8\,V \\
\hline
\end{tabular}
\end{center}
\label{spstnd}
\end{table}%

Finally, we built spectral templates by combining some of these spectra and convolving them with
an appropriate rotational profile (25 km\,s$^{-1}$) and scaled by a factor of 0.5 in order to make them consistent with that
of V1082\,Sgr, whose lines are reduced by the light contribution from the companion.
The final templates were T4660 (HD 131977), T4900 (average of HD 160346 and HD 23356),
T5080 (HD 192310, HD 22049), T5250 (HD 149661, HD 165341), and T5430 (HD 69830, HD 152391).

We looked for metallic lines that can be used for spectral-type classification in the spectral
regions less contaminated by the companion.
Fig.~\ref{sptype} shows the spectrum  of V1082\,Sgr in comparison with the reference spectra.
Some \ion{V}{1}-to-\ion{Fe}{1} and \ion{Cr}{1}-to-\ion{Fe}{1} line ratios are sensitive to temperature.
Another good indicator is the aspect of the $\lambda$4227 \ion{Ca}{1} line.
In all cases, the line ratios suggest a spectral type of K2\,V (templates T4900 to T5250).

The surface of the donor  is likely to have  spots, similar to the chromospherically active stars \citep{2005LRSP....2....8B}, 
so the temperature  measurements vary with  orbital phase \citep[e.g.][]{2007AN....328..813W}. 
It is  clear from the spectrum that the donor star in \sgr\  is not a giant star, but  whether the spectrum
corresponds exactly  to a normal-size main-sequence star or is slightly larger is hard to tell from the spectral features. We are inclined 
to identify the donor star of \sgr\ as K1\,V-K2\,V.
The high-resolution spectra thus confirm the results obtained previously by \citet{2016ApJ...819...75T}, but
make the classification more reliable.

\section{Rotation}
\label{rotation}

%The donor star rotation has to be synchronized with the orbital motion, since the 
The {\sl Kepler K2} mission 80 days of continuous photometry shows that 
the orbital period is present in the light curve during the low state (Paper I).  The period appears as a smooth, nearly sinusoidal, 
single-humped wave during a deep minimum when the optical flux is produced predominantly by the K star. 
The  observed light curve is interpreted as an evidence  that the K star is not deformed by overflowing its Roche lobe 
but has a spot or spots  on its surface, reflecting its rotational period.  
%We have numerous evidences, including the spectra obtained in this
%study, which show that  the optical emission in deep minima comes predominantly from the K star.
Since the rotation can be safely assumed to be synchronous, the projected rotational velocity can give information 
about the radius of the late-type stellar companion. In a first trial, we  determine $v\sin i$ using the method developed by  
\citet{2011A&A...531A.143D}, which allows us to deal with the line blending present in late-type stars. 
This technique reconstructs 
the rotational profile from the cross-correlation function of the object spectrum against a sharp-lined template and derives 
the rotational velocity from the first zero of the Fourier transform (FT) of the rotational profile. In these calculations, we adopted limb-darkening 
coefficients from \citet{2013A&A...556A..86N}.%\footnote{The adopted values were 0.575 for aperture 14, 0.496 for ap. 8, and 0.483 for ap. 7.}.

Measuring the rotational velocity is a difficult task for this object. Several factors conspire against a reliable determination.  
%The rotational broadening is relatively low in comparison with the spectral resolution. 
\citet{2011A&A...531A.143D} mentioned that the method 
works satisfactorily when the rotational broadening is larger than the instrumental profile (or other broadening effects) by a factor of 2. 
%In this object we are close to that limit: \fed{the full-width half-maximum (FWHM) of the instrumental profile is about 18-20 \kms,
%while the rotational profile has FWHM$\sim 37$\kms.}.\\
%\fed{{\em Alternatively we can write :}\\
In the present case, the rotational broadening ($v\sin i\approx 27$ \kms) is relatively low in comparison to the spectral resolution 
(the FWHM of the instrumental profile is about 18-20 \kms).
On the other hand, the S/N ratio is modest ($\sim 20$ for individual observations, $\sim$80 
for the average spectrum), which is aggravated by the dilution of the spectral lines due to the light contribution of the companion. 
%Also some problems with focus of the spectra across the CCD exist, 
%\fed{\sout{altough this problem is mitigated by the reference spectra being taken 
%in the same conditions as the object, both being affected in the same way.}\\
%{\em I remove (move forward) this sentence because this is not mitigated when using the method by Diaz et al}}\\

\begin{figure}[t]
\centering
\includegraphics[height=8.6cm, width=6cm, angle=-90]{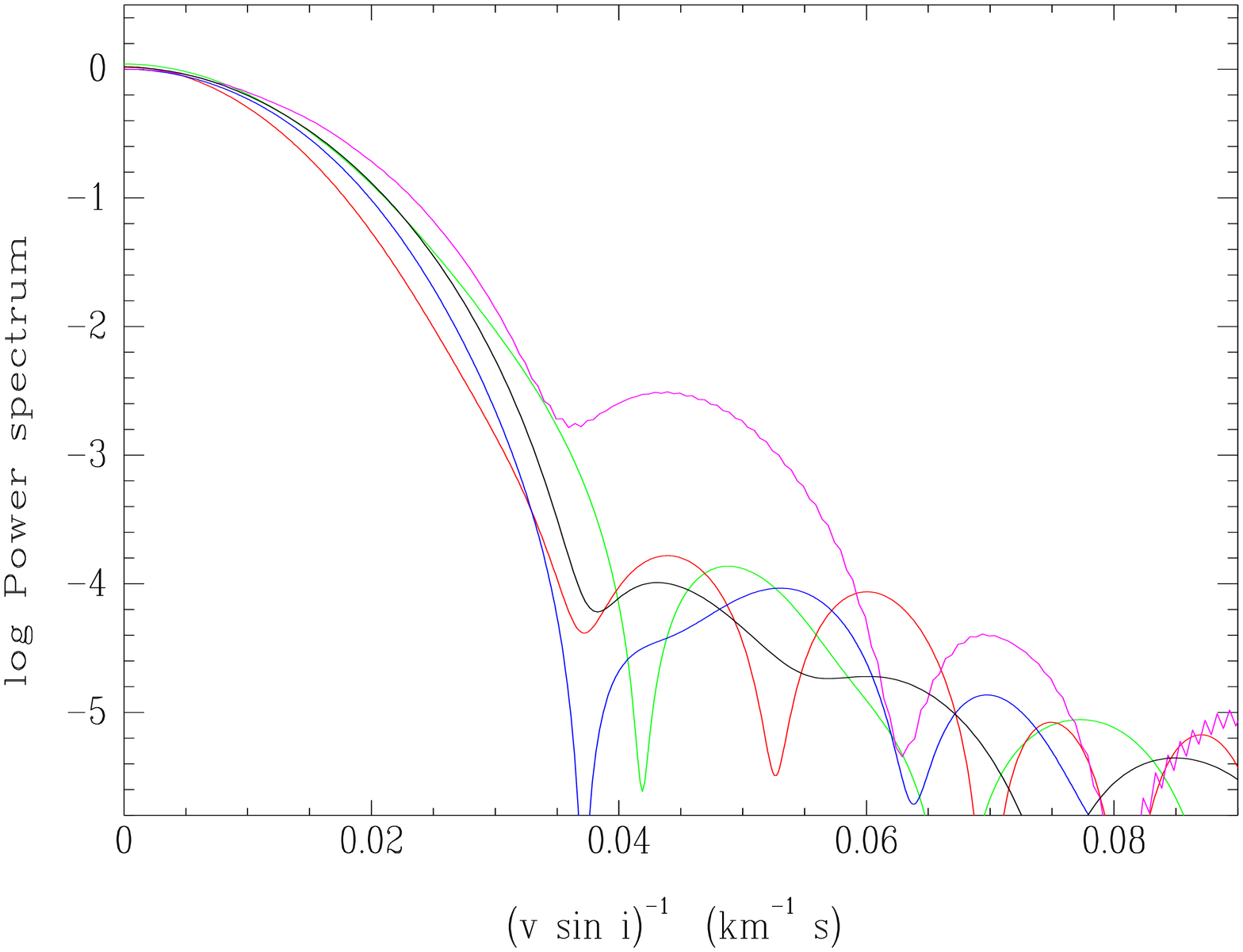}
\includegraphics[height=8.6cm, width=6cm, angle=-90]{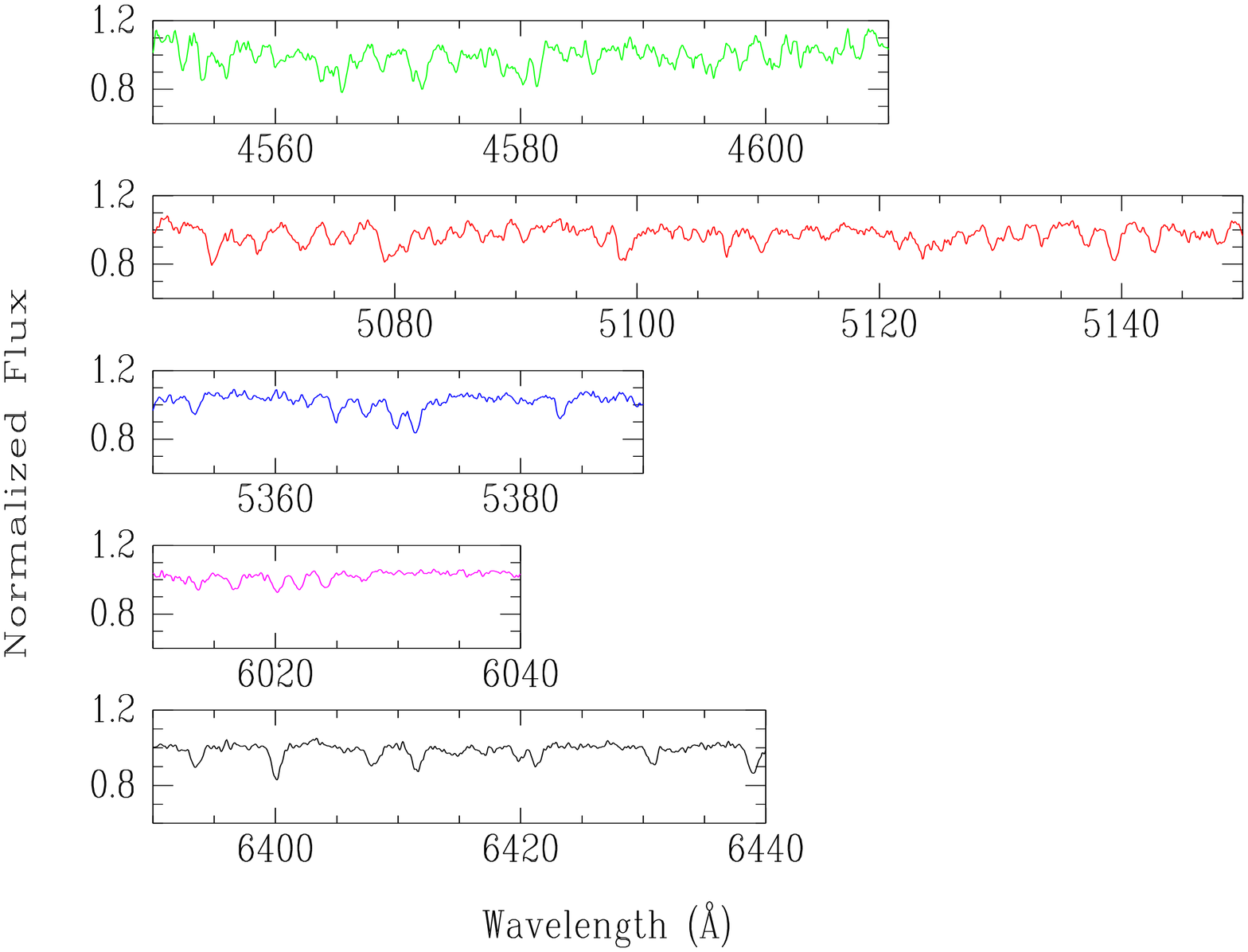}
\caption{The FT of the autocorrelation function between the selected regions containing absorption lines and the template is presented in the top panel.
The first zero of the FT corresponds to the rotational velocity (see \citet{2011A&A...531A.143D} for details). 
The selected spectral regions are plotted in the bottom five panels. The FT power of each region is plotted with a corresponding color. }\label{pow}
\end{figure}

Finally, the useful spectral regions are rather limited, since regions contaminated by the companion emission lines must be excluded, 
as well as those containing the strongest lines of the late-type star itself, since such line profiles are affected by pressure broadening.
Some examples of the power spectra of the line profiles are shown in Fig.~\ref{pow}.

%\begin{figure}[t]
%\centering
%\includegraphics[height=8.5cm, width=8cm]{power.eps}
%\caption{Comparison of the spectral morphology of V1082\,Sgr with reference spectra of different spectral types.}\label{pow}
%\end{figure}

We applied the mentioned technique to a few selected regions presented in the bottom part of Figure\,\ref{pow} .
In the best three regions, we obtained $v\sin i$ = 27.7$\pm$1.0 \kms, 27.6$\pm$ 2.5 \kms, and 26.3$\pm$1.5 \kms, 
while some other failed to define a reliable zero FT at all. 

As a second strategy, we estimated $v\sin i$ by comparing the target spectrum with a template previously convolved with different rotational profiles.
We used as a reference the observed spectrum of HD\,182488, which was convolved with rotational profiles between 22 and 32 \kms.  
The comparison was done by cross-correlating each spectrum against the original reference spectrum and measuring the FWHM of 
the central peak of the cross-correlation function. The rotation of reference star is below the spectral resolution and has a negligible contribution to the peak FWHM. 

The main advantage of this method with respect to the former is that the contribution of the instrumental profile is less  
critical.  Particularly, the mentioned problem of deficient focus over the CCD (variable instrumental profile along the spectrum) is largely mitigated 
by the fact that both  object and  reference spectra are taken under the same conditions and, therefore, are affected in the same way. 
Hence, we considered this strategy more reliable and applied it to 10 spectral regions ranging from
%by the fact that the reference spectra being taken in the same conditions as the object, both being affected in the same way
%\footnote{You may wonder why to use the Fourier technique. 
%The Fourier technique would give much better results in faster rotators, where the line profile differs largely from a Gaussian, making the second strategy 
%not very good in such case.
%Hence we consider  this second strategy more reliable  and applied it to ten small spectral regions} ranging from 
4250  to 6450\AA. 
We used  as a template the spectrum of HD\,182448 (a sharp-lined K0V star) taken with the same instrument.  It is probably slightly hotter 
than the object, and their chemical abundances do not match exactly. However, the very low $v\sin i = 0.6$\,\kms\  rotational velocity of the template 
and variety of lines used for measurements make it 
suitable for the task. 

We also considered the effect of velocity smearing on our measurements as an outcome  of relatively long exposure times (1200\,s) used in observations. 
We convolved  synthetic spectrum representing 
the intrinsic stellar spectrum (with $v\sin i = 26.5 $\,\kms) with boxy kernels of widths equal to the RV variation during 
the exposure time. We calculated the mean spectrum and measured the rotation with the same procedure as described in this section. 
The results showed that our measurement of $v\sin i$ would be about 0.3\,\kms\  above the true value, that is, well below the uncertainties cited below.

%and a high S/N spectrum built combining several  spectra of K2\,V-type stars from the ELODIE database \citep{2001A&A...369.1048P}.  
The obtained values of $v\sin i$  average  at $25.3\pm2.4$\,\kms\  within small error spread. However, for three regions below $\lambda4900\AA$, 
the values are slightly smaller and show tendency to decrease toward shorter wavelengths. The contribution of the donor star at shorter 
wavelengths drops rapidly when the object is not in  deep minima, hence the measurements at the blue end of the spectra are less reliable. 
Excluding these three values, we obtain $v\sin i = 26.5\pm1.4$\,\kms\ of a very stable subsample.

Considering the possible differences in the instrumental profile between the observations of the object and reference star, we adopted a value 
of 26.5$\pm 2.0$\,kms  for the projected rotational velocity of the donor companion. Either method has its uncertainties and limitations, 
and probably the statistically derived error bars are underestimated, but  matching results of independent measurements assure that the  
result is realistic.

\section{Stellar Parameters}
\label{param}

If we assume that the rotation of the K-type star (marked with subindex "d'' for "donor") is synchronized with the orbital motion, the relative radius of this star can be written as
\begin{equation}
\frac{R_{\mathrm {d}}}{a}=\frac{q}{(1+q)} \cdot \frac{v_{\mathrm d}\sin i}{K_{\mathrm {d}}},\label{eq0}
\end{equation}
where $q=M_{\mathrm {wd}}/M_{\mathrm {d}}$ is the mass ratio.
On the other hand, according to \citet{1983ApJ...268..368E}, the effective radius of the Roche lobe in units of the orbital semiaxis can be calculated through the expression
\begin{equation}
\frac{R_\mathrm{L}}{a} = \frac{0.49}{0.6 + q^{2/3}\cdot \ln\left(1+q^{-1/3}\right)},\label{eq1}
\end{equation}
 which is better that  1\% for any value of the mass ratio\footnote{In the original formula by Eggleton, we have changed $q$ to $q^ {-1}$, 
 since in his work, the mass ratio is $M_{\mathrm d}/M_{\mathrm {wd}}$.}.

 From these two equations, we obtain the ratio between the stellar radius and the critical radius:
 \begin{equation}
\frac{R_{\mathrm d}}{R_\mathrm{L}} = \left(\frac{v_\mathrm{d}\sin i}{K_{\mathrm d}}\right)\cdot \frac{\left[0.6 + q^{2/3} \ln\left(1+q^{-1/3}\right)\right]}{0.49\,(1+q^{-1})}
\end{equation}   

The first factor is known from the spectroscopy: $(v_\mathrm{d}\sin i/K_{\mathrm d}) = 0.586\pm0.045$. 
Then, the constraint  imposed by the Roche lobe on the donor star volume ($R_{\mathrm d} \leq R_{\mathrm L}$), 
provides an upper limit for the mass ratio:  $q \leq 1.42\pm0.2$.
%\fed{\sout{The donor star  would be, therefore, less massive than its compact companion.}}
%Assuming that the companion is a white-dwarf, its mass is expected to be in the range 0.5--1.4 M$_\odot$, and hence the donor should have $M_1=$0.35--0.99 M$_\odot$.
% However, the primary might be smaller than the Roche lobe, so the mentioned $q$ values is in fact  an upper limit.

Fortunately, we have  strong constraints on the mass of the compact companion from the X-ray spectroscopy.
\citet{2013MNRAS.435.2822B} derived a WD mass of $M_{\mathrm {wd}}=0.64\pm 0.04$\,\msun\ 
by  modeling the spectrum obtained by {\sl XMM-Newton} EPIC and {\sl Swift} BAT.

For a comprehensive  analysis of the possible configurations, we calculated $a$, $M_{\mathrm d}$, $R_{\mathrm d}$ and $R_\mathrm{L}$  for 
different possible values of $q$.
More precisely, from $M_{\mathrm {wd}}$ and $q$,  we calculated the mass of the donor star $M_{\mathrm d}$ and the total mass, and from the {\sl Kepler} equation, 
we calculated the orbital semiaxis.
Then the radii of the donor $R_\mathrm{d}$  and that of its Roche lobe $R_\mathrm{L}$ were calculated 
from equations~\ref{eq0} and \ref{eq1}. 

\begin{figure}[t]
\centering
\includegraphics[width=5.cm,bb= 140 160 480 580]{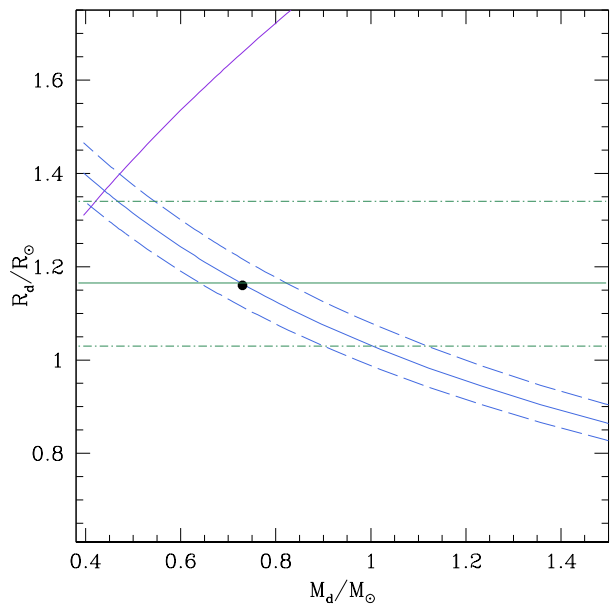}
\caption{Mass-radius diagram  for the donor star in \sgr. The black dot represents our best solution. 
The blue line (with limits marked as dashed lines) is the solution from the dynamical constraints. The violet line is the Roche lobe size.
The green line is the radius of the donor star determined by the distance modulus The limits mark an error strip related to the temperature 
ambiguity (dash-dotted lines). See text for the explanation.}\label{fig:param}
\end{figure}

The results are shown in Fig.~\ref{fig:param}.
The radius of the donor star $R_{\mathrm d}$ is plotted as a function of $M_{\mathrm d}$ with a blue line; blue dashed lines mark the uncertainty interval of the radius due to the error in $v\sin i$.
 The radius of the Roche lobe corresponding to the donor star is plotted with a violet line.
% Black lines are isochrones  for logarithmic ages (from right to left) 8.0, 9.0, 9.2, 9.4, 9.6, 9.8, and 10.0.
%We have used PARSEC stellar models \citep{2012MNRAS.427..127B} for solar metallicity.
%In the same model grids we interpolated isotherms for $T_\mathrm{eff}$ = 4900, 5000, 5100, 5200 (red lines, from left to right). 
%This temperature range correspond approximately to the spectral type of the donor star according to 
%Mamajec's calibration\footnote{\texttt{http://www.pas.rochester.edu/$\sim$emamajek/EEM\_dwarf\_UB VIJHK\_colors\_Teff.txt}}.
%\footnote{\em Tell me if you agree  with the adopted temperature range. The Mamajec calibration is: K0V=	5280, K1V=5170, K2V=5040, K3V= 4840.} }

%\fed{\sout{With a stringent constrain on the WD mass we can deduce all the basic parameters of these binary system we get the $q=0.78$ 
%and estimate the mass of the donor 
%star to be about 0.82\,\msun. This would correspond to  a main-sequence K2 star, which squares perfectly with the spectral classification.}} 
The adopted solution is marked in Fig.~\ref{fig:param} by a black dot.
The full solution of the binary system parameters are summarized in Table\,\ref{tab:param2}. 
The consigned uncertainties have been calculated taking into account the observational errors in $P_\mathrm{orb}$, $v_\mathrm{d}\sin i$, $M_\mathrm{wd}$, and $K_\mathrm{d}$, and the error of $M_\mathrm{d}$ which has been estimated from the spectral type (see Fig.~\ref{fig:param}).

\begin{table}[h]
\caption{Deduced absolute parameters of \sgr.}
\centering
\vspace{2 mm}
\begin{tabular}{lcr@{$\pm$}l}  
\hline \hline
Parameter & Units &\multicolumn{2}{c}{Value} \\
\hline
M$_{\mathrm {d}}$  & \msun & 0.73 & 0.04 \\
M$_{\mathrm {wd}}$ & \msun & 0.64 & 0.04 \\
$i$                &$\deg$ & 23.3 & 1.3  \\
$q$                &       & 0.88 & 0.09 \\
$a$                & \rsun & 4.25 & 0.07 \\
R$_{\mathrm {d}}$  & \rsun & 1.16 & 0.11 \\
R$_{\mathrm {L}}$  & \rsun & 1.66 & 0.05 \\
\hline
\end{tabular}
\label{tab:param2}
\end{table}

Thus, the donor  fills only a fraction of its Roche lobe (about one-third of the corresponding volume), 
 although it is significantly ($\sim$70\%) larger than a nonevolved main-sequence star of the same mass.
 In fact, the evolution of stars of such masses is so slow that this mass-radius relation can be explained only
 as the result of a nonstandard evolution.
 
 \subsection{Stellar parameters: Gaia Distance}

While this paper was in the final stages of preparation, the second Gaia data release  \citep[Gaia DR2;][]{2016A&A...595A...1G},
which provides precise  parallaxes for an unprecedented number of sources, became available.  According to Gaia DR2,
the distance to \sgr\ is $669\pm13$\,pc. This allows us to impose direct and stringent limits on the size of the donor star. 
Using the faintest visual magnitude V=14.8, repeatedly recorded during deep minima in over 1500 days; 
interstellar reddening of E(B-V)=0.15 \citep{1998ApJ...500..525S},   and T$_{\mathrm {eff}}=4930$\,K, 
a temperature corresponding to a K2-type star we obtain R$_{\mathrm {d}}=1.165$\,\rsun. If we take into account the uncertainty in temperature 
of up to  $350$\,K, which is the largest factor affecting the luminosity,   we obtain a range of values of the donor star R$_{\mathrm {d}}=1.03 - 1.34$\,\rsun,
independent of the mass of the binary components. This range of values is marked in Fig.~\ref{fig:param} as a  horizontal strip within  dash-dotteded green lines, 
with the central value exactly corresponding to what we deduced before the distance was known.

%The radius/mass relation for the donor is 1.13/0.82= 1.38 (solar units), while for standard stellar models 
%\citep{2012A&A...541A..41M}} the ratio R/M is $\sim0.89$ at ZAMS
%$\sim1.20$ when H is exhausted at the star center (the bluest point of the track in the HR diagram).

\section{Analysis of Emission Lines}
\label{emlines}

\begin{figure}[t]
\centering
\includegraphics[width=9.0cm, bb = 30 150 620 690, clip=]{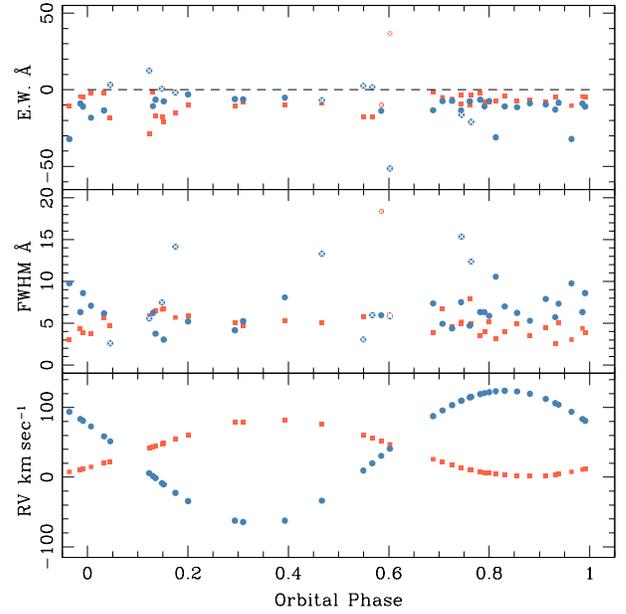}
\caption{Parameters of two components of emissionline \ion{He}{2} marked by red squares and blue circles. 
The RVs are presented in the bottom panel, the FWHMs in the middle panel, and EWs  in the top panel.    
Three iterations of the deblending procedure 
%using IRAF-{\sl splot} 
were made to find solutions that correspond to two components with strictly periodical RVs. 
The central wavelengths, FWHMs, and core intensities of these components were used 
to produce the profiles shown in Figure 7.
%Figure\,\ref{profiles}. 
The horizontal dotted line in the top panel corresponds to zero. Points above that line are invalid and marked
with crosses.  Also marked with crosses are the points with extremely broad components, as can be seen in the middle panel.   }
\label{rvhe2}
\end{figure}

\begin{figure*}[t]
\centering
\includegraphics[width=8.0cm, bb = 30 150 620 690, clip=]{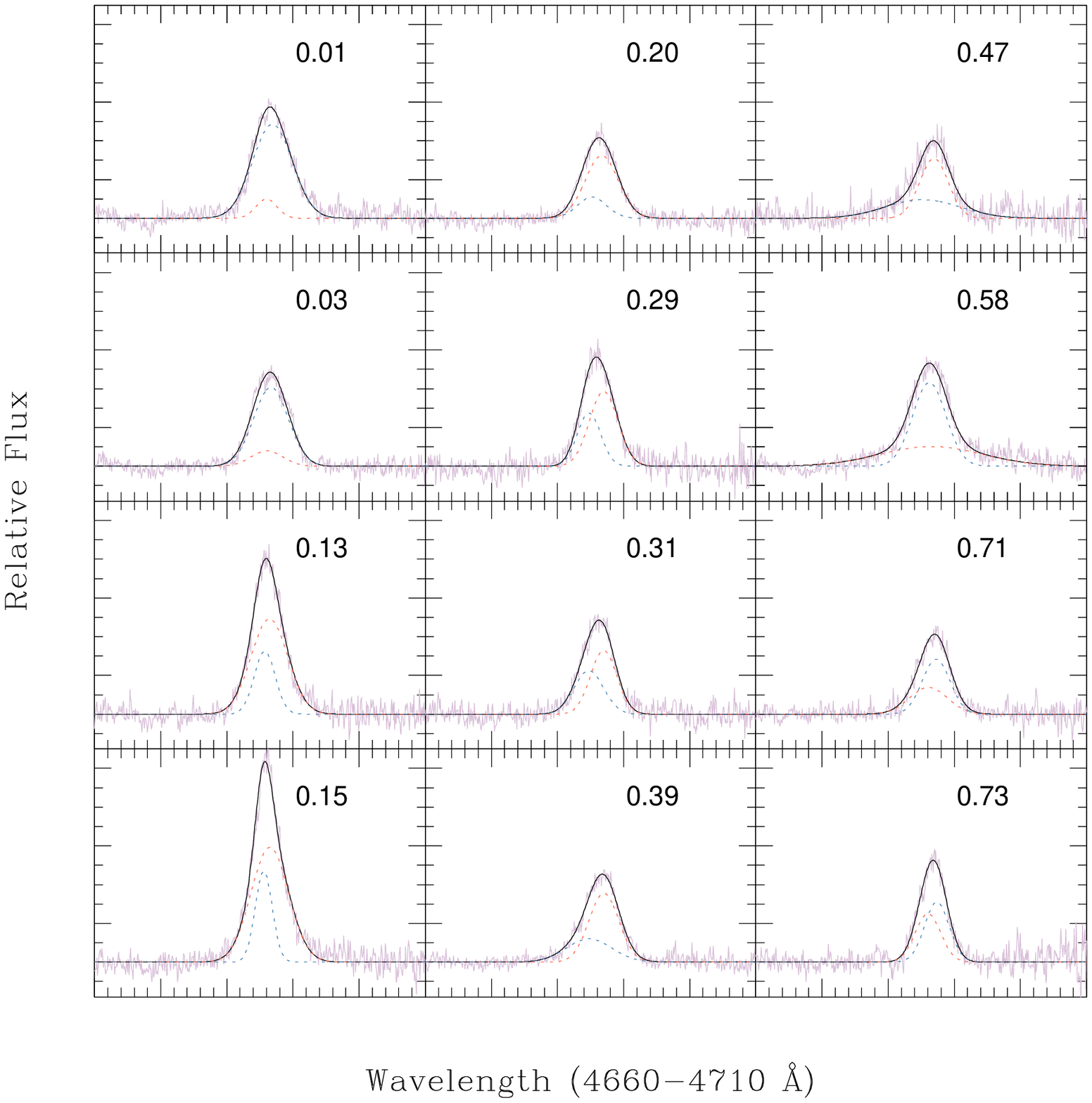}
\hspace{-1.2cm}
\includegraphics[width=8.0cm, bb = 30 150 620 690, clip=]{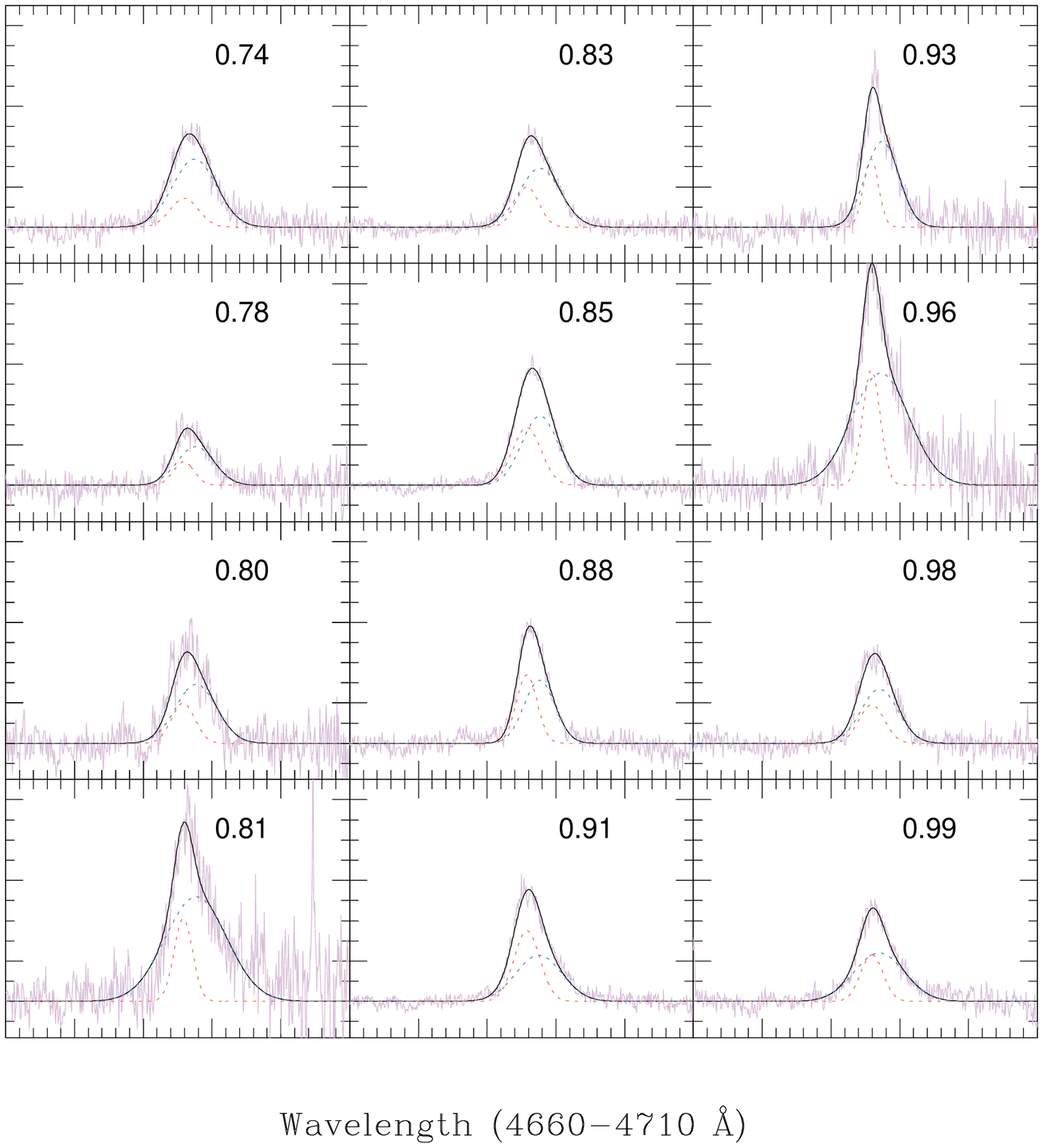}
\caption{Examples of disentangling emission-line \ion{He}{2}  into  two components, plotted by red  and blue dotted lines.  
Black curves correspond to the sum of the two components, 
while the observed spectra are plotted in light violet. The vertical axes are in normalized counts, with a continuum 
corresponding to one unit. The numbers in each box indicate the orbital phase at which that spectrum was taken.
}
\label{profiles}
\end{figure*}

From the previous lower-resolution spectroscopy, it was known that the emission lines are single-peaked and  
fairly symmetric  \citep{2016ApJ...819...75T}. However, measuring emission lines in the low-resolution spectra as a single line 
has been producing  RVs of low amplitude and a huge scatter of points that was not the result of measurement errors. 
The amplitudes and phases  of different emission lines were vastly different too, indicating that the situation was not assessed correctly.  
%The  RV measurements were not producing expected  sinusoidal curves.
%The individual measurements of RV show large
%deviations from the sinusoidal fit to the data. Calculated phases and amplitudes vary from one line to another. 
%The high-resolution spectroscopy did not improve 
%the situation when the lines are measured by a single Gaussian, or Lorenzian. 
%The amplitude of RV curves turns 
%out less than 20\,\kms, the phase displays significant shifts from line to line. 

At higher resolution, the profiles of lines appear to be not as symmetric as was thought, and we made an attempt to discern  
them into two components. The attempt was particularly successful 
in the case of the \ion{He}{2} $\lambda4685.75\,\AA$\ line, which is somewhat sharper than the Balmer lines. We first 
used the deblending option of the IRAF {\sl splot} procedure 
to fit each profile with two Gaussians. For each spectrum, we obtained a pair of RV values corresponding to two components. 
By eye inspection, we selected one component per spectrum  that appeared to belong to a sinusoidal pattern in the RV-vs-orbital phase diagram.
It was not possible to identify the right component in every spectrum, but in more than half of the cases, the selected points formed a clear pattern
that could be  fit with a sine curve  with the orbital  period.  We calculated the central wavelength 
of that component for each spectrum and iterated  the deblending procedure for all spectra, but this time keeping 
one component's central wavelength fixed and another as free. 
We also set the FWHMs of both Gaussians as free parameters. The measurements of the second 
component formed a periodic pattern too, reasonably well fitted with 
another sinusoid. Finally, we did a third iteration of deblending by fixing the central wavelengths of both components according 
to the calculated sinusoids and leaving  the FWHM and intensities variable.  We obtained excellent fits to the line profiles in most of the cases. 
In only in 10 spectra  out of 38,  the program could not find  two emission components with reasonable parameters. 

In order to separate emission lines  into two components, we assumed that they move sinusoidally. We defined the most readily identifiable 
component and  then refined the parameters of the other one. 
Of course, this is a very idealistic approach; the reality has to be much more complicated, 
because the gas streams  producing emission lines have intrinsic velocities and directions not coinciding with the orbital motion.

The RV measurements of these two components
are plotted against the orbital phases  in the bottom panel of Figure\,\ref{rvhe2}. The points correspond to the calculated 
central wavelengths that were fixed in the last deblending attempt; hence, they form a perfect sinusoids.  The orbital 
phases are calculated according to the ephemeris determined by absorption spectra. In the middle panel of   Figure\,\ref{rvhe2} the 
FWHMs of both components are plotted. Most are very consistent with the general trend. Some strongly deviate, indicating problems
usually related to the phases in which both components crisscross  and become indistinguishable.  
In such a case, the program tends to pick a broad second component  extending to the noise in the continuum or use an 
absorption component. Such cases appear as points above the dotted line corresponding to zero in the top  
panel of   Figure\,\ref{rvhe2}, which depicts the equivalent weights (EWs) of the components. Those points in the plot are marked by red crosses.
Among them are also some points that have large FWHM, inconsistent with the average.  

In Figure\,\ref{profiles} examples of line profile fits are presented. We omitted orbital phases (marked by numbers in each panel) 
in which  the lines were not split into two components correctly, with one component  being in absorption.  In the other case, one of the 
components is  getting too wide, like in phases  $\phi=0.47$ and 0.58,  
as a consequence of both of them getting too close and  difficulty in separating them.
Although the procedure of selecting  components and fixing central wavelengths  is somewhat arbitrary, the final result is encouraging. 
Both components show  larger RV amplitudes than the  entire line  when measured with a single Gaussian.

The component  marked in red color in 
Figures\,\ref{rvhe2} and \ref{profiles} 
maintains rather stable FWHM throughout all orbital phases.   Neither component  appears to vary strictly in antiphase 
from the absorption line. Neither it is expected to be related to the stellar elements of the binary. Invalid points constitute a quarter 
of all measurements  and do not influence the general interpretation.

We assume that the emission lines are produced by the ionized gas between the stellar components. The lines practically disappear 
when the accretion is halted and the system is in 
a deep minimum. Hence, we do not have RV measurements for the first 3 nights, when the lines were faint or absent. 
After the accretion is reestablished, the 
disentangled components  of  the \ion{He}{2} line act similar to polars, showing phase shifts and high velocities related to the stream 
intrinsic velocity rather than the orbital velocity.
The brightness of the object varies significantly in very short timescales (Paper I), and the intensity of the lines changes accordingly.

\begin{figure}[t]
\centering
\includegraphics[width=8cm, bb = 80 100 700 700,clip=]{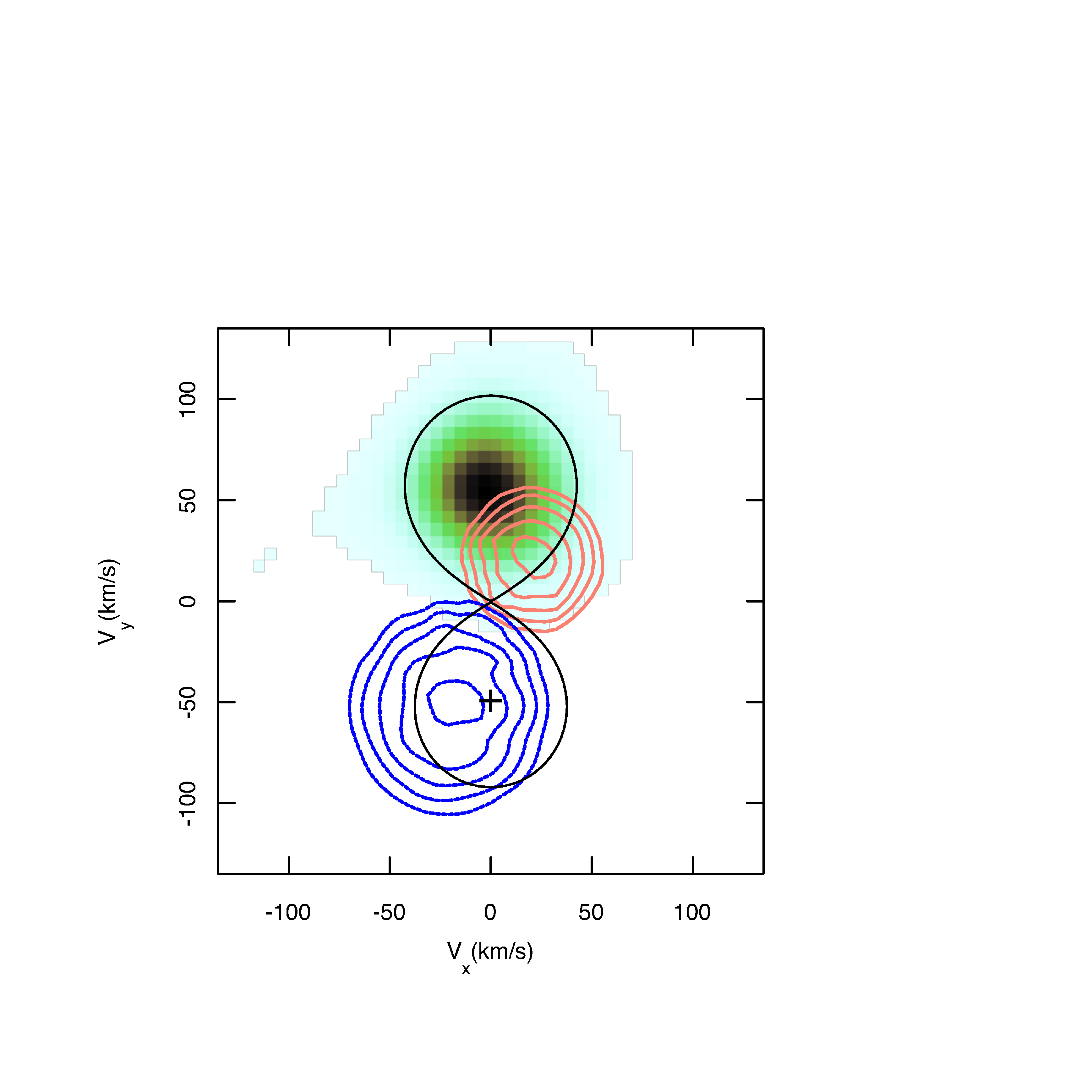}
\caption{Doppler maps of  an absorption line and two components  of \ion{He}{2} emission.
All lines are artificially made with RVs of measured lines and average FWHM and intensities. 
The image map  corresponds to the donor star while the higher-velocity (blue)
and  the lower-velocity (red) components are presented in the form of isochrones. The Roche lobes 
and stellar position are  marked according to Table\,\ref{tab:param2}.}
\label{dopmap}
\end{figure}

In an attempt to better understand the components of the emission lines, we used Doppler tomography. 
The system has a rather small  inclination  for sensible
tomograms to be made. Observed lines or their components of variable intensity and width make the reading of tomograms impossible. 
Hence, we made  artificial lines of  fixed FWHM and intensity with RVs and phases
of the real components. The FWHM was fixed at 5 and 7.5\,\AA\  for 
the lower- and higher-velocity  components, respectively, 
according to their averages. 
We also added an artificial narrow (FWHM$=0.7$\,\AA\  absorption line emanating from the secondary to make the Doppler map more
illustrative. Doppler maps of these three lines are presented in Figure\,\ref{dopmap}. 
The  inclination angle and masses derived from observations  and listed in  Table\,\ref{tab:param2} 
were used to plot the Roche lobes and star locations on the Doppler map.

The intensity of the lines is not irrelevant, but since there is no a well-established pattern of line intensity  and FWHM change 
with the phase, we kept them constant in all cases. The artificial spectra were distributed unevenly by phases, according to 
observations. Hence,  the  spots on the map are not ideally round.  
One has to bear in mind that this image is simplistic and may not reflect all the complexity of ionized gas 
distribution in the binary.

The Doppler map  illustrates  where in the velocity coordinates the lines
originate.  The high-velocity emission component is concentrated in the third quadrant. It may  correspond to the accretion 
curtain component observed in polars  
as described by \citet[][see Figure 5]{1999MNRAS.304..145H}. The ballistic component is obviously absent; instead, 
we have a lower-velocity component in the first quadrant, practically in between the stars.
This component may originate in the gas flowing along the coupled magnetic lines (magnetic bottleneck). 
%We see no evidence of heated/irradiated donor star  from the component separation of \ion{He}{2} line.  

\citet{2013MNRAS.436..241P} detected similar  line components (corresponding to the  magnetic bottleneck)
in a detached binary, but with an M-type donor star.  The RV and phasing of the fainter of two components
of H$_\alpha$ that they observed indicate that it comes from  gas located in between the stars. 
%The other emission line component in their case comes from the heated side of M-star itself. 
%The mass transfer and accretion rates are much
%smaller in WD+M detached binaries and there is no significant X-ray emission. The cooling of accreted material
%in such systems happens through the cyclotron emission.  But the binary is compact and the 
%WD is hot enough to irradiate the M dwarf. }

%In \sgr\ the lower excitation emission lines  (H$_\alpha$ and H$_\beta$ to a smaller extent) show a narrow emission lines from the donor star 
%when the system is in the deep minimum, which we consider  to be a result of chromospheric activity \citep{2016ApJ...819...75T}
%rather than irradiation effect.  
To pinpoint  the donor star on the tomogram, we use the simulated absorption line. The absorption line
produces a spot in the tomogram corresponding to the center of mass of the K star.

\section{Discussion and Conclusions}
\label{conclud}

We conducted high-resolution spectroscopy of \sgr, proposed to be one of two possible candidates for a pre-polar with an early-K-type 
donor star  \citep{2017ASPC..509..489T}. This spectroscopic study was accompanied by a prior continuous, 80 day 
photometry of the object by the {\sl Kepler K2} mission.  The results of the
photometry are reported in a separate publication (Paper I). 
For a wholeness of argument, we must repeat the main conclusion of Paper I: the orbital modulation detected in 
the light curve of \sgr\ persists during the deep minima, when  the donor star is the predominant source of light. 
It is argued that the variability is caused by  spots on the surface (either hot,  cold, or a combination of both). No double hump is detected in the light curve.  
Hence, it is concluded  that the donor star is not ellipsoidal  and that it rotates with the orbital period.
%This is a significant result which confirms our hypothesis that \sgr\ is a 
%detached binary. 
By measuring the rotational velocity of the donor star, we can go a step further and estimate other important 
parameters of the binary system. 

These parameters are listed in Table\,\ref{tab:param2}. According to them, the donor star occupies about 70\% of its corresponding 
Roche lobe radius, nevertheless exceeding the main-sequence star size of similar temperature. 
This means that the donor star has probably departed from the  
zero-age main sequence. 
%and probably follows one of the tracks plotted as (red) curves in Figure\,\ref{fig:param}. 

These estimates and conclusions are based on the assumption that the WD in this detached binary accretes as a polar and has a rather average mass
for a WD according to the spectrum of the  X-ray emission. 
%Frankly, we see no alternative to this assumption. 
%Apart from the 
%X-ray spectrum indicating appropriate temperature and emissivity area, high-excitation lines like \ion{He}{2} can only be  explained by
%an energetic ionization source.

While there is no direct evidence of an MWD, it is obvious that the donor star is magnetically active and has large spot(s) on its 
surface. A possible explanation of the mass transfer and accretion on the WD is the magnetic coupling, capture of the stellar wind 
from the donor star, and channeling (siphoning) of the matter onto the magnetic pole of the WD. 
Doppler maps  of the \ion{He}{2} line, which consists of two components, testify that the  ionized  gas  is not emitting  from 
areas and spots associated  with the accretion disk or  streams proper to ordinary polars.
This has been observed already in  
compact binaries comprised of an MWD and an M-type star \citep{2013MNRAS.436..241P} for which the pre-polar phenomenon is well established. 
However,  significant differences may 
appear if the binary contains an earlier spectral-type donor star.

These differences are related to the fact that a  more massive donor star has a large convective envelope with an intense chromospheric activity, 
has a higher-rate stellar wind, and is prone to evolution in a Hubble time, unlike M-dwarf companions.
The separation of the binary is also larger, and the mass ratio is reversed (as compared to pre-polars with M components or CVs), 
making the Roche lobe of the donor star 
much bigger than the size of a main-sequence star and hence providing a long time for evolution until the binary becomes 
semidetached.    

There is a lack of knowledge regarding stellar winds in general and their dependence on rotational velocity due to the large spread in rotation rates 
of isolated stars at young ages. However, the latest models predict that the mass-loss rate due to the stellar wind depends 
moderately on the mass of low-mass stars, and more significantly 
on the rotational velocity \citep[][and references therein]{2015A&A...577A..27J}. Most of the observational studies
of mass loss by stars of solar mass and below presume that the  rotation slows down because of magnetic braking and 
the diminishing stellar wind. \citet{2002ApJ...574..412W} estimated that at younger ages, the solar wind may have been 
as much as $10^3$ times stronger.  This enables a steep increase of the mass-loss rate estimate from the  fast-rotating donor star 
in \sgr\  compared to identical single stars. Chromospheric activity and a larger surface area 
can further  fuel the mass loss.  The mass-loss rate appears to depend on the X-ray surface flux as a power 
law \citep{2002ApJ...574..412W}. 

The X-ray flux from \sgr\ in the minimum is probably due to the  donor 
star, while the accretion on the WD is halted. The flux matches an upper limit observed from similar magnetically active stars.
Thus, all prerequisites exist to expect a mass accretion rate a few orders higher than that in pre-polars with an M-star companion.

Another thorny issue is how a mass loss from the donor star converts into the mass accretion on the WD. 
\citet{2012ApJ...746L...3C} demonstrated that matter lost through the stellar wind will not always find its way
to the WD. There are configurations in which all the wind can effectively siphon to the MWD. 
Curiously, such configurations require  antialigned and require rather modest magnetic fields. However, it is still 
incomprehensible  how the accretion rate reaches an estimated $2\times10^{-9}$\,\msun\, yr$^{-1}$\ 
\citep{2013MNRAS.435.2822B} in \sgr.  With the Gaia distance, this rate is 1.3 times less, but is still of the same order. 

The accretion is not continuous, however. It is regularly interrupted by deep minima
states where there is no evidence of accretion at all. We speculate that the cessation of accretion is related to the 
broken magnetic coupling. The deep minima last only a few orbital periods, then the accretion is restored.

Such a high accretion rate exceeding that of an ordinary CVs is unusual and raises questions about how it will affect the evolution of the binary
system.  The excess nitrogen abundance probably indicates that \sgr\ has formed from  massive progenitors and underwent an episode
of thermal time-scale mass transfer, as was pointed out by \citet[][and references therein]{2013MNRAS.435.2822B}.

We found independent observational  evidence  suggesting that  \sgr\ is a detached binary containing a slightly evolved early-K star. 
If this proposition is correct, we see no alternative 
to the magnetic coupling as the means of transferring matter and angular momentum from the donor to an MWD.  This object
corroborates the existence of pre-polars  with earlier spectral-type donor stars. It offers an explanation as to why magnetic 
systems are not found in the search for detached WD+FGK detached binaries \citep{2016MNRAS.463.2125P,2017MNRAS.472.4193R}. 
Yet the object presents certain challenges. It is necessary to estimate the magnetic field strength of the WD and its temperature and 
mass directly by UV observations. Furthermore, it is necessary to calculate possible evolution scenarios in conditions of a high mass transfer rate
while the system is still detached and where it will evolve.

\acknowledgments
This paper has made use of data obtained from the ESO Science Archive Facility under request numbers 360705, 360734, 361117, and 360635.
This work has made use of data from the European Space Agency (ESA)
mission {\it Gaia} (\url{https://www.cosmos.esa.int/gaia}), processed by
the {\it Gaia} Data Processing and Analysis Consortium (DPAC). Funding
for the DPAC has been provided by national institutions, in particular
the institutions participating in the {\it Gaia} Multilateral Agreement.
DGB is grateful to CONACyT for grants allowing his postgraduate studies. 
M-SH thanks  CONICYT-PFCHA/Doctorado Nacional/ 2017-21170070, and JVHS is 
supported by a Vidi grant awarded to N. Degenaar by the Netherlands Organization for Scientific Research (NWO).
GT and SZ acknowledge PAPIIT grants IN108316/IN-100617 and CONACyT grant 166376. 

\facility{OAN SPM} %, \facility{PROMPT}, \facility{SWIFT}

\bibliography{paperII}

\end{document}